\documentclass[10pt,letter]{emulateapj} 

\begin{document}
\title{On weak and strong magnetohydrodynamic turbulence}
\author{Jean Carlos Perez}
\author{Stanislav Boldyrev}
\affiliation{Department of Physics, University of Wisconsin at Madison, 1150 University Ave, 
Madison, WI 53706, USA; {\sf jcperez@wisc.edu, boldyrev@wisc.edu}}

\begin{abstract}
Recent numerical and observational studies contain conflicting reports on 
the spectrum of magnetohydrodynamic turbulence. In an attempt to clarify the 
issue we investigate anisotropic incompressible magnetohydrodynamic
turbulence with a strong guide field~$B_0$. We perform numerical
simulations of the reduced MHD equations in a special setting that
allows us to elucidate the transition  between weak and strong
turbulent regimes. Denote $k_{\|}$, $k_\perp$ characteristic
field-parallel and field-perpendicular wavenumbers of the
fluctuations, and $b_{\lambda}$ the fluctuating field at the scale
$\lambda\sim 1/k_{\perp}$. We find that when the critical balance
condition, $k_{\|}B_0\sim k_{\perp} b_{\lambda}$, is satisfied, the 
turbulence is strong, and the energy spectrum is $E(k_{\perp})\propto
k^{-3/2}_{\perp}$. As the $k_{\|}$ width of the spectrum increases,
the turbulence rapidly becomes weaker, and in the limit $k_{\|}B_0\gg
k_{\perp} b_{\lambda}$, the spectrum approaches $E(k_{\perp})\propto
k_{\perp}^{-2}$.  The observed sensitivity of the spectrum to the
balance of linear and nonlinear interactions may explain the
conflicting numerical and observational findings where this balance
condition is not well controlled.  

%\pacs{95.30.Qd, 52.30.Cv}
\end{abstract}

\maketitle

\section{Introduction}
In this Letter we investigate homogeneous and
steadily driven incompressible magnetohydrodynamic (MHD)
turbulence. In practical applications, such as fusion devices, solar
wind, and interstellar medium, turbulence is driven by various
large-scale instabilities, and turbulent energy is then spread over a
broad range of spatial scales due to nonlinear interactions until
small dissipative scales are reached where the energy is removed from
the system. In the interval of scales between the injection and
dissipation regions turbulence
properties are thought to be universal~\citep[e.g.,][]{frisch,biskamp}. 

The MHD equations describing the evolution of magnetic and
velocity fluctuations ${\bf b}(\bf x,t)$ and ${\bf v}(\bf x,t)$ in the
presence of a guide field ${\bf B_0}$ can be represented in the
so-called Els\"asser variables ${\bf z}^{\pm}={\bf v}\pm {\bf b}$:   
\begin{eqnarray}
\partial_t {\bf z}^{\pm} \mp ({\bf V}_A\cdot \nabla){\bf
  z}^{\pm}+({\bf z}^{\mp}\cdot \nabla){\bf z}^{\pm}=-\nabla P +{\bf 
  f}, 
\label{mhd1} 
\end{eqnarray}
where ${\bf V}_{A}={\bf B}_{0}/\sqrt{4\pi \rho}$ is the Alfv\'en
velocity,  $\rho$ is the fluid density, $P$ is the pressure that is
determined from the incompressibility condition, $\nabla \cdot {\bf
  z}^{\pm}=0$, ${\bf f}$ represents a large-scale forcing, and we omit the
terms representing small viscosity and resistivity. 
The linear term on the left-hand side of equations~(\ref{mhd1}), 
$({\bf V}_A\cdot \nabla){\bf z}^{\pm}$,
is responsible for advection of the $z^+$ and $z^-$ wave
packets, with the Alfv\'en velocity along the guide field. The nonlinear   
term, $({\bf z}^{\mp}\cdot \nabla){\bf z}^{\pm}$, describes the 
interaction of turbulent fluctuations, and it is responsible
for the energy transfer among different spatial scales.  The nonlinear
term is considered small if    
\begin{eqnarray}
k_{\|}B_0\gg k_{\perp} b_{\lambda},
\label{weak-turb}
\end{eqnarray} 
where $k_{\|}$ and $k_{\perp}$ are typical field-parallel and
field-perpendicular wavenumbers of the fluctuations' spectrum,  
and $b_{\lambda}$~($\ll B_0$) is the typical magnitude of fluctuations at the
scale~$\lambda\sim 1/k_{\perp}$. This regime is referred to 
as ``weak turbulence.''  

The regime when the nonlinear term in not formally   
small will be called ``strong turbulence.''  One can argue
that in strong turbulence the following critical balance condition
should be maintained at all scales~\citep{goldreich}: 
\begin{eqnarray}
k_{\|}B_0\sim  k_{\perp} b_{\lambda}. 
\label{crit}
\end{eqnarray}
Indeed, during the characteristic time of nonlinear interaction,
$\tau_N\sim 1/(k_{\perp} b_{\lambda})$, the fluctuations   
become correlated along the guide field up to a distance 
$l_\|\sim V_A\tau_N$. This causality condition ensures the critical
balance~(\ref{crit}). Depending on the way turbulence is excited, it
satisfies either condition~(\ref{weak-turb}) or~(\ref{crit}) in a
certain range of scales.

Recent numerical simulations and analytic modeling suggest that in the
case of strong turbulence (\ref{crit}), the field-perpendicular 
energy spectrum is $E(k_{\perp})\propto k_{\perp}^{-3/2}$
\citep{maron,muller,boldyrev,boldyrev2,mason,mason2}.  
However, geophysical and astrophysical observations often exhibit  
somewhat steeper spectra~\citep[e.g.,][]{goldstein,bale}. 
This raises the question of to what extent such systems can be described 
in the framework of MHD turbulence.

In the present work we conduct direct numerical simulations of 
reduced MHD equations, driven by a force with varying 
$k_{\|}$~spectral width. This provides a unifying numerical setting 
allowing one to address the regimes of weak and strong turbulence 
in the same framework. We observe that when the critical balance~(\ref{crit}) 
is satisfied, 
the spectrum of strong MHD turbulence is close to~$-3/2$. When the critical balance 
condition~(\ref{crit}) is
even slightly broken, the spectrum steepens. As the
weak turbulence condition~(\ref{weak-turb}) becomes better satisfied, 
the spectral exponent approaches~$-2$ in accord with the
theory of weak turbulence \citep{ng96,galtier}. The observed sensitivity of the spectrum to the 
forcing details may explain conflicting results of numerical and 
astrophysical observations, where the spectral properties of forcing are either 
not well controlled or not well known. 

According to the standard derivation~\citep[e.g.,][]{biskamp}, the reduced MHD equations  
are valid in the region $k_{\perp} \gg k_\|$; 
therefore, their applicability in the strong turbulence regime~(\ref{crit}) is justified.  
Their applicability in the weak turbulence regime~(\ref{weak-turb}), 
however, requires an explanation, which we provide in the following sections.

\section{Weak MHD Turbulence.} 
When the   
condition~(\ref{weak-turb}) is satisfied, one may assume that
turbulence consists of shear-Alfv\'en and pseudo-Alfv\'en  
waves, weakly interacting with each other. In the absence of 
nonlinear interaction, the waves would have random phases, and  
the Gaussian rule could be applied to express their higher order  
correlation functions  through the second-order ones.\footnote{Many 
  papers contributed over the years to the development of fundamental
  ideas on MHD turbulence, see e.g., the reviews 
  in~\citep{biskamp,ng}. The general methods of weak turbulence theory
  are reviewed in~\citep{zakharov,newell}.}  \citet{galtier}
developed a perturbative theory of weak MHD turbulence based on such
random phase approximation. By expanding the MHD eq.~(\ref{mhd1}) up 
to the second order in the nonlinear interaction and using the
Gaussian rule to split the fourth-order correlators, they  
derived a closed system of kinetic equations governing 
the wave energy spectra.

These equations demonstrate that wave energy cascades in the Fourier 
space in the direction of large~$k_{\perp}$, and the universal spectrum 
of wave turbulence is established in the region~$k_{\perp}\gg k_{\|}$. In 
this limit the dynamics of the shear-Alfv\'en waves decouple 
from the dynamics of the pseudo-Alfv\'en 
waves, and the pseudo-Alfv\'en waves are passively scattered by the
shear-Alfv\'en ones.  The kinetic equation for the energy spectrum of shear-Alfv\'en waves, 
$e({\bf k}, t)$, derived by \citet{galtier}, then reads:
\begin{eqnarray}
\partial_t e({\bf k})= \int M_{{\bf k}, {\bf p} {\bf q}}
e({\bf q}) [e({\bf p})-e({\bf k})]\delta(q_\|)
d_{\bf k,pq}
\label{galtier-eq}
\end{eqnarray}
In this expression, the interaction 
kernel is $M_{{\bf k}, {\bf p}{\bf q}}=({\pi}/{V_A}){({\bf
    k}_{\perp}\times{\bf q}_{\perp})^2({\bf k}_{\perp}\cdot {\bf
    p}_{\perp})^2}/({k_{\perp}^2 q_{\perp}^2 p_{\perp}^2})$,  
and we adopt the shorthand notation $d_{\bf k,pq}  
\equiv \delta({\bf k}-{\bf p}-{\bf q})\,d^3p\,d^3q $. The phase-volume  
compensated energy spectrum is then  
calculated as~$E({\bf k}, t)dk_{\|}\,dk_{\perp}= e({\bf k},
t)k_{\perp}dk_{\|}\,dk_{\perp}$.   
The stationary 
solution of equation~(\ref{galtier-eq}) was found 
analytically and verified numerically in \citep{galtier}. 
It has the general form~$E({\bf k})=f(k_{\|})k_{\perp}^{-2}$, where
$f(k_{\|})$ is an arbitrary function that is smooth at~$k_{\|}=0$.

It should be noted, however, that the derivation of~(\ref{galtier-eq}) based
on the weak interaction approximation is not rigorous. 
As follows  from Eq.~(\ref{galtier-eq}),   
only the $q_\|=0$~components of the energy spectrum $e({\bf q})$ are
responsible for the energy transfer. However, if we apply Eq.~(\ref{galtier-eq}) to
these dynamically important  
components themselves, that is, if we set $k_{\|}=0$ in (\ref{galtier-eq}), 
we observe an inconsistency. Indeed, the perturbative approach implies
that the linear  
frequencies of the waves are much larger than the 
frequency of 
their nonlinear  
interaction. The nonlinear interaction 
in~(\ref{galtier-eq}) remains nonzero as $k_{\|}\to 0$ while the
linear frequency of  
the corresponding Alfv\'en modes, $\omega_k=k_{\|}V_A$, vanishes. Therefore, as shown 
by~\citet{galtier}, the additional assumption of smoothness of the 
function $f(k_\|)$ at $k_\| =  0$ is crucial for deriving  
the spectrum $E({\bf k}) \propto k_{\perp}^{-2}$.

A definitive numerical verification of such a spectrum  
seems therefore desirable.  While numerical integration of a scattering 
model based on MHD equations expanded up to the second order in 
nonlinear interaction does reproduce the~$-2$ exponent~\citep{bhattacharjee-ng,ng}, 
this spectrum has not yet been 
confirmed in direct numerical simulations of systems~(\ref{mhd1}). The major problem 
faced by such simulations  
is to simultaneously satisfy the two conditions,  
$k_\perp\gg k_{\|}$ and $k_{\|}B_0\gg k_\perp b_{\lambda}$, which is hard to achieve
with present-day computing power. In the next section we discuss a numerical setting in which the 
spectrum of weak turbulence can be verified.

\section{Model equations.}
An important fact concerning
Eq.~(\ref{galtier-eq}) was emphasized   
by~\citet{galtier-chandran}. They noted that there exists a
dynamical system  
that leads to exactly the same kinetic equation~(\ref{galtier-eq}) in
the weak turbulence regime~(\ref{weak-turb}),  
{\em without} any additional restrictions on~$k_{\perp}$ and~$k_{\|}$.   
To derive this system, we note that in the universal regime where Eq.~(\ref{galtier-eq}) 
is applicable, the polarization vectors of the pseudo-Alfv\'en modes are almost parallel to the 
guide field. One can therefore consider a system where such modes are eliminated   
for {\em arbitrary}~${\bf k}$ by restricting the initial MHD system to    
field-perpendicular fluctuations,~${\tilde {\bf z}}^{\pm}$:
\begin{eqnarray}   
\displaystyle \partial_t {\tilde {\bf z}}^{\pm} \mp ({\bf V}_A\cdot \nabla){\tilde
    {\bf z}}^{\pm}+({\tilde {\bf z}}^{\mp}\cdot \nabla){\tilde {\bf
      z}}^{\pm}&=&-\nabla_{\perp} P 
+\frac 1{Re}\nabla^2{\tilde {\bf      z}}^{\pm}
+{\tilde {\bf f}},\label{restricted} \nonumber\\ 
 \nabla \cdot {\tilde {\bf z}}^{\pm}&=&0. 
\end{eqnarray}
The fluctuating fields here have only two vector components, 
${\tilde {\bf z}}^{\pm}=\{{\tilde z}^{\pm}_1, {\tilde z}^{\pm}_2, 0 \}
$, but depend on all three spatial coordinates. Although 
system~(\ref{restricted}) is not presented in 
\citep{galtier-chandran}, their analysis of~(\ref{mhd1})  
is equivalent to solving such a system. Formally,  
system~(\ref{restricted}) is equivalent to the  
reduced MHD equations~\citep[e.g.,][]{biskamp,Shebalin}. 
The principal difference is in the limits of validity:  the reduced MHD model 
is applicable only for~$k_{\perp}\gg k_{\|}$, 
while we consider system~(\ref{restricted}) without any restrictions. 

Within the formalism of the weak turbulence theory 
both systems (\ref{mhd1}) and~(\ref{restricted}) lead to the same kinetic 
equation~(\ref{galtier-eq}) for the shear-Alfv\'en turbulence. 
We thus conclude that the wave energy spectrum 
obtained from the full MHD system~(\ref{mhd1}) under the 
assumption $k_{\perp}\gg k_{\|}$  should coincide with the spectrum  
obtained from the restricted system~(\ref{restricted}), where 
the condition $k_\perp \gg k_\|$ is not required. On the other hand,
strong turbulence is expected to develop when  
$k_\|B_0\sim k_\perp b_\lambda$, which is precisely the domain in 
which reduced MHD provides a good approximation of the full MHD model.    
This opens an effective way for numerical investigation of both strong and 
weak MHD turbulence in the same framework, which is the goal of 
the present Letter.

\section{Numerical method and results.}
We solve numerically the 
restricted MHD model~(\ref{restricted}) using a fully 
dealiased pseudo-spectral technique in a periodic box that is
elongated along the guide field ${\bf B}_0$ with aspect ratio
$2\pi:2\pi:L_z$. The random force ${\tilde {\bf f}}$ has no component along
$z$, it is solenoidal in the $x-y$ plane and its Fourier coefficients
outside the range $1 \leq k_{\perp} \leq 2$, $(2\pi/L_z) \leq k_\| \leq
(2\pi/L_z)n_z$ are zero, where integer $n_z$ determines the width of
the force spectrum in $k_\|$. The Fourier coefficients inside that
range are 
Gaussian random numbers with amplitude chosen so that the resulting
rms velocity fluctuations are of order unity.  
The individual random values are refreshed independently at time
intervals $\tau= 0.1~L_\perp/(2\pi U_{rms})$. The parameters~$n_z$ and $L_z$ control  
the degree to which condition~(\ref{weak-turb}) or~(\ref{crit})
is satisfied at the forcing scale. Note that we  do {\em not} drive the $k_\|=0$~modes  
but allow them to be generated by nonlinear interactions. 
The Reynolds number is defined as $Re=U_{rms}(L_{\perp}/2\pi)/\nu$, where 
$L_{\perp}$~$(=2\pi)$ is the field-perpendicular box size, $\nu$ is fluid
viscosity, and $U_{rms}$~$(\sim 1)$ is the rms value of velocity
fluctuations. In our case magnetic resistivity and fluid viscosity are  
the same, $\nu=\eta$. 
The system is evolved until a stationary state is reached,
as determined by the time evolution of the total energy of the
fluctuations. A typical run produces from 30 to 60 snapshots. 
The field-perpendicular energy spectrum is obtained by 
averaging the angle-integrated Fourier spectrum, $E(k_\perp)=0.5\langle|{\bf v}({\bf
k_\perp})|^2\rangle k_\perp+0.5\langle|{\bf b}({\bf
k_\perp})|^2\rangle k_\perp$, over field-perpendicular planes 
in all snapshots. 

\begin{figure}[tbp]
\includegraphics[width=0.41\textwidth]{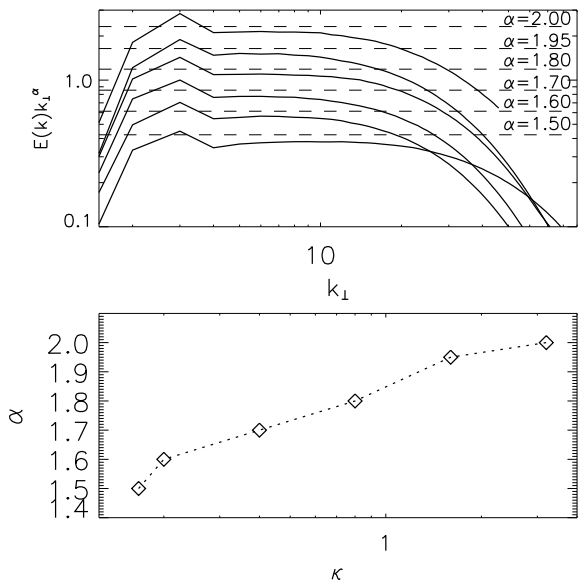}
\caption{The field-perpendicular energy spectrum~$E(k_{\perp})$ of MHD
  turbulence,
calculated by direct numerical integration of~(\ref{restricted}). The
  presented cases
correspond, from bottom to top, to
  $\kappa=1/6,\,1/5,\,2/5,\,4/5,\,8/5,\,16/5$. For clarity,
the curves in the top panel are
arbitrarily offset in the vertical direction.}
\label{spectrum}
%\vskip-3mm                                                                                                                                                          
\end{figure}

We performed a series of simulations for $B_0=5$, $L_z=5L_\perp$ and $n_z=1,\,2,\,4,\,8,\,16$.   
We used the resolution $256^3$ mesh points in these simulations (and $Re=800$), 
except for the case $n_z=16$, where the resolution 
was $512\times 512\times 256$ (and $Re_\perp =2000$).  We also performed a simulation 
with $B_0=5$, $L_z=6L_\perp$, and $n_z=1$ and the resolution $512\times 512\times 256$ (and $Re_\perp =2000$). 
Fig.~\ref{spectrum} shows the field-perpendicular energy 
spectra  for each run.  All the runs have different 
values of parameter $\kappa\equiv (2\pi/L_z)n_z$ that measures deviation from the 
critical balance~(\ref{crit}) condition. 

\begin{figure}[tbp]
\includegraphics[width=.41\textwidth]{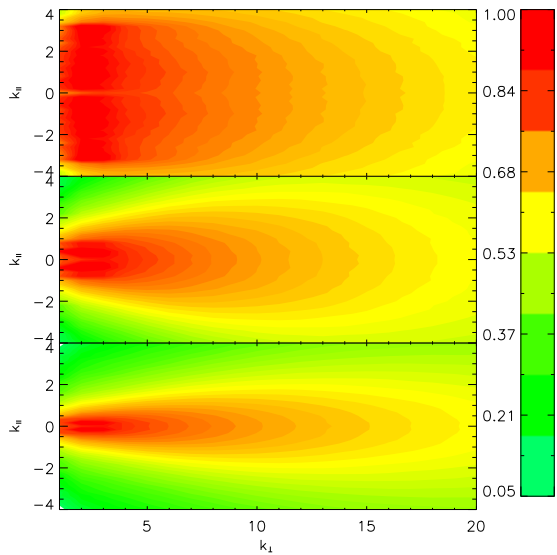}
\vspace*{-0.5cm}
\caption{Contour plots of the anisotropic energy spectra
for the cases $\kappa=1/6$~(bottom), $\kappa=4/5$~(center),  and
$\kappa=16/5$~(top). The colors represent energy in a log
scale, normalized to the energy of the most dominant large-scale
mode.}
\label{spectrum2d}
\end{figure}

We found that as the spectral width of the forcing along
$k_\|$ increases, higher and higher frequency modes of the velocity and 
magnetic fields are  
excited. For run $\kappa=1/6$, all the forced modes have linear frequency
$\omega = k_\|B_0 \approx 1$, which corresponds to a critically balanced
forcing.\footnote{It is important to keep in mind that the critical
  balance condition should be interpreted in a order-of-magnitude
  fashion, and that its validity is ultimately verified by the
  resulting spectrum of strong turbulence.}  In this case, the spectrum is $E(k_\perp)\propto 
k_\perp^{-3/2}$.  This result is consistent with recent 
numerical simulation of full MHD by~\citet{mason2}, since 
the reduced MHD system approximates the full MHD system when 
the critical balance condition is satisfied.   
As we increase the parameter~$\kappa$, we break   
the critical balance condition at the forcing scales. As a result, the spectrum 
monotonically steepens from $-3/2$ in the strong turbulence 
case to $-2$ in the weak turbulence case, as shown in Fig.~\ref{spectrum}. 

Fig.~\ref{spectrum2d} shows isocontours of the full energy spectrum $E(k_{\|}, k_{\perp})$ 
as a function of~$k_\perp$ and~$k_\|$ for the three typical 
cases $\kappa=1/6,\, 4/5, \, 16/5$.  The bottom frame presents the energy
distribution for the case $\kappa=1/6$, where the random force 
preserves the critical balance. As the cascade continues deeper 
into the inertial range, higher 
frequency modes~$\omega = k_\|B_0$ are generated by virtue of
nonlinear interactions, just enough to maintain the critical balance 
condition at all scales, and establish a strong turbulence
spectrum. As the frequency of the forced Alfv\'en modes increases 
in the cases $\kappa =(2, \dots , 16)/5$, the parallel cascade is slightly
inhibited as the weaker interaction among the large scale Alfv\'en modes 
dominates the energy transfer to smaller scales, resulting in a 
steepening of the field-perpendicular energy spectrum. This can be 
seen in the middle and top frames in
Fig.~\ref{spectrum2d}, where the distribution of energy becomes more
and more elongated along $k_\perp$ rather than $k_\|$.

\section{Discussion.}
Our numerical results demonstrate that if 
the energy spectrum has a limited extent, increasing 
the~$k_\|$ width of the forcing spectrum leads to the energy spectrum of 
weak turbulence~$\propto k_{\perp}^{-2}$. If however  
the $k_\|$ width of the forcing is limited, but we can achieve arbitrarily high 
resolution in the $k_\perp$ direction, the interaction between Alfv\'en modes 
will eventually become 
strong enough to satisfy critical balance and establish a strong 
turbulence spectrum.  

This is partly supported by the following derivation. It can be proved that   
turbulent fluctuations described by system~(\ref{restricted}) satisfy
the exact relation:
\begin{eqnarray}
\langle \delta {\tilde z}^{\pm}_{l}(\delta {\tilde {\bf
  z}}^{\mp})^2
\rangle=-2\epsilon^{\mp} r_{\perp}, 
\label{pp-restricted}
\end{eqnarray} 
where $\delta {\tilde {\bf z}}^{\pm}= {\tilde {\bf z}}^{\pm}({\bf
  x}+{\bf r}_{\perp})-{\tilde {\bf z}}^{\pm}({\bf x})$, and 
$\delta {\tilde z}^{\pm}_{l}=\delta {\tilde {\bf z}}^{\pm}\cdot
{\bf r}_{\perp}/r_{\perp}$ is the longitudinal 
component of $\delta {\tilde {\bf z}}^{\pm}$. Averaging is taken over the
statistical ensemble, or over time and spatial position~{\bf x}.   
In this formula, $\epsilon^{\pm}$ are the constant rates of~${\tilde z}^{+}$
and~${\tilde z}^{-}$ energy dissipation. In the isotropic case, that is, without the guide field, 
the relation analogous to~(\ref{pp-restricted}) was derived by~\citet{politano-pouquet}.  
We now prove the following general inequality: 
\begin{eqnarray}
\langle \delta {\tilde z}^{\pm}_{l}(\delta {\tilde {\bf
  z}}^{\mp})^2
\rangle^2 \leq\langle (\delta {\tilde {\bf z}}^{\pm})^2 \rangle
\langle (\delta {\tilde {\bf z}}^{\mp})^4\rangle . 
\label{mhd-inequality-restricted}
\end{eqnarray}
The first step is to use the Schwartz inequality, $\langle \delta {\tilde z}^{\pm}_{l}(\delta {\tilde {\bf
  z}}^{\mp})^2 \rangle^2\leq \langle (\delta {\tilde z}_{l}^{\pm})^2 \rangle
\langle (\delta {\tilde {\bf z}}^{\mp})^4\rangle $; the second
step is to note that~$(\delta {\tilde z}^{\pm}_{l})^2\leq (\delta {\tilde {\bf
  z}}^{\pm})^2$, which completes the proof. 

Both sides in the expression~(\ref{mhd-inequality-restricted}) have finite limits
as viscosity and resistivity go to zero. In the inertial interval, the 
correlation functions on the right-hand side of this expression have
a power-law behavior, that is, $\langle (\delta {\tilde {\bf z}}^{\pm})^2\rangle
\propto r_{\perp}^{{\zeta}_2}$ and $\langle (\delta {\tilde {\bf
  z}}^{\pm})^4\rangle \propto r_{\perp}^{{\zeta}_4}$ (we assume statistical symmetry 
between ${\tilde {\bf z}}^+$ and ${\tilde {\bf z}}^-$).  Since the left
hand side of (\ref{mhd-inequality-restricted}) scales as  $\propto r_{\perp}^2$, and the
inequality should hold for arbitrarily small~$r_{\perp}$, we obtain the exact
result 
\begin{eqnarray}
\zeta_2+\zeta_4\leq 2. 
\label{bound-mhd}
\end{eqnarray}

This inequality is useful for evaluation of these 
exponents from numerical simulations or experiments since the {\em
  ratio} of these exponents is well measured by the method
of extended self-similarity~\citep{benzi}. The inequality~(\ref{bound-mhd})   
then provides a boundary on the turbulence energy spectrum that  
is related to the second-order scaling exponent as   
$E(k_{\perp})\propto k_{\perp}^{-1-\zeta_2}.$  
In our case, the scaling exponent ${\zeta}_4$ is usually close to
$2{\zeta}_2$ within small intermittency
corrections,  which can be checked numerically~\citep[e.g.,][]{muller-biskamp-grappin}.  
Inequality~(\ref{bound-mhd}) 
then implies that $3{\zeta}_2\leq 2$, and,
therefore, the field-perpendicular energy spectrum cannot be essentially  
steeper than $E(k_{\perp})\propto k^{-5/3}_{\perp}$ in the limit~$k_\perp \to \infty$. 

Note, however, that in our numerical findings the spectral exponent~$-5/3$ is   
not distinguished in any way; rather, the field-perpendicular energy 
spectrum of strong MHD turbulence is flatter and closer to~$-3/2$. 
This is consistent with recent results of~\citet{muller,mason2} and also with 
high-resolution simulations of isotropic MHD turbulence by~\citet{haugen,mininni}. 
Astrophysical observations of the solar wind and of the interstellar medium reveal  
the presence of MHD turbulence, and find support for both $-5/3$ and $-3/2$ spectral 
exponents~\citep[e.g.,][]{goldstein,goldstein-roberts,bale,borovsky,podesta,smirnova}. 
However, statistics of such data are often not good enough to distinguish 
between ``$-5/3$'' and ``$-3/2$'' with confidence. On the numerical side, 
simulations of MHD turbulence in the framework of reduced MHD were performed in many  
works~\citep[e.g.,][]{dmitruk,gomez,rapazzo}; however, either the 
simulation domain was not anisotropic to ensure the critical balance condition~(\ref{crit}), 
or the driving force was not spatially homogeneous, for example, applied at the boundary 
of the domain. 

Our results suggest that the interpretation 
of observational and numerical results may be obscured 
if the $k_\|$ and $k_{\perp}$ structure of the spectrum is either not well 
measured or not well controlled, in which case it is hard to deduce whether 
the field-parallel dynamics have been  captured and whether the universal regime of 
MHD turbulence has been established.

\acknowledgments
This work was supported by the US 
Department of Energy under grant DE-FG02-07ER54932, and 
by the NSF Center for Magnetic
Self-Organization in Laboratory and Astrophysical Plasmas
at the University of Wisconsin-Madison.  
High-performance-computing resources were provided by the Texas
Advanced Computing Center (TACC) at the University of Texas at Austin
under the NSF-Teragrid Project TG-PHY070027T. URL:http://www.tacc.utexas.edu

\clearpage

%\begin{figure}[tbp]
%\includegraphics[width=1\textwidth]{f1.eps}
%\caption{The field-perpendicular energy spectrum~$E(k_{\perp})$ of MHD turbulence, 
%calculated by direct numerical integration of~(\ref{restricted}). The presented cases 
%correspond, from bottom to top, to $\kappa=1/6,\,1/5,\,2/5,\,4/5,\,8/5,\,16/5$. For clarity, 
%the curves in the top panel are 
%arbitrarily offset in the vertical direction.}
%\label{spectrum}
%%\vskip-3mm
%\end{figure} 

%\begin{figure}[tbp] 
%\includegraphics[width=.7\textwidth]{f2.eps}
%\caption{Contour plots of the anisotropic energy spectra 
%for the cases $\kappa=1/6$~(bottom), $\kappa=4/5$~(center),  and
%$\kappa=16/5$~(top). The colors represent energy in a log
%scale, normalized to the energy of the most dominant large-scale mode.}    
%\label{spectrum2d}
%\end{figure}


\begin{thebibliography}{99}
\bibitem[Bale, et al(2005)]{bale} Bale, S. D., Kellog, P. J., Mozer, F. S., Hornbury, T. S., \& Reme, H., 
Phys. Rev. Lett., {\bf 94} (2005) 215002.
\bibitem[Benzi et al.(1993)]{benzi} Benzi, R., Ciliberto, S.,
   Tripiccione, R., Baudet, C., Massaioli, F., and Succi, S. (1993)
   Phys. Rev. E {\bf 48} R29.
\bibitem[Bhattacharjee \& Ng(2001)]{bhattacharjee-ng} Bhattacharjee, A. \& Ng, C. S., ApJ {\bf 548} 
(2001) 318. 
\bibitem[Biskamp(2003)]{biskamp} Biskamp, D., 2003, {\em Magnetohydrodynamic Turbulence.} (Cambridge University Press, %%@
Cambridge).
\bibitem[Boldyrev(2005)]{boldyrev} Boldyrev, S., ApJ {\bf 626} (2005) L37.
\bibitem[Boldyrev(2006)]{boldyrev2} Boldyrev, S., Phys. Rev. Lett. {\bf 96} (2006) 115002. 
\bibitem[Borovsky(2006)]{borovsky} Borovsky, J., J. Geophys. Res. (2006) submitted. 
\bibitem[Dmitruk et al.(2003)]{dmitruk} Dmitruk, P., Gomez, D., Matthaeus, W., Phys. Plasmas, {\bf 10} (2003) 3584. 
\bibitem[Frisch (1995)]{frisch} Frisch, U., 1995, {\em Turbulence}. (Cambridge University Press, Cambridge).
\bibitem[Galtier et al.(2000)]{galtier} Galtier, S., Nazarenko, S. V., Newell, A. C., \& Pouquet, A., 
J.~Plasma Physics {\bf 63} (2000) 447.
%\bibitem[Galtier et al.(2002)]{galtier2} Galtier, S., Nazarenko, S. V., Newell, A. C., \& Pouquet, A., 
%ApJ {\bf 564} (2002) L49.
\bibitem[Galtier \& Chandran(2006)]{galtier-chandran} Galtier, S. \& Chandran, B.D.G., Phys. 
Plasmas, {\bf 13} 114505 (2006).
\bibitem[Goldreich \& Sridhar(1995)]{goldreich} Goldreich, P. \& Sridhar, S., ApJ {\bf 438} (1995) 763.
\bibitem[Goldstein \& Roberts(1999)]{goldstein-roberts}	Goldstein, M. L.; Roberts, D. A., Phys. Plasmas, {\bf 6} (1999) 4154.
\bibitem[Goldstein et al.(1995)]{goldstein} Goldstein, D. A. Roberts, D. A., \& Matthaeus, W. H., 
Annu. Rev. Astron. Astrophys. {\bf 33} (1995) 283. 
\bibitem[Gomez et al.(2005)]{gomez} Gomez, D. O., Mininni, P.D. \& Dmitruk, P., Physica Scripta T~{\bf 116} (2005) 123.   
\bibitem[Haugen et al.(2003)]{haugen} Haugen, N.E.L., Brandenburg, A. \& Dobler, W., ApJ~{\bf 597} (2003) L141.  
\bibitem[Maron \& Goldreich(2001)]{maron} Maron, J., \& Goldreich, P., ApJ {\bf 554} (2001) 1175.
\bibitem[Mason et al. (2006)]{mason} Mason, J., Cattaneo, F., \& Boldyrev, S., Phys. Rev. Lett, {\bf 97} (2006) 255002. 
\bibitem[Mason et al. (2007)]{mason2} Mason, J., Cattaneo, F., \& Boldyrev, S., arXiv:0706.2003 (2007).
%\bibitem[Boldyrev, Mason \& Cattaneo(2006)]{boldyrev3} Boldyrev, S., Mason, J. \& Cattaneo, F., (2006) arXiv:astro-ph/0605233. 
\bibitem[Mininni \& Pouquet(2007)]{mininni} Mininni P.D. \& Pouquet, A., arXiv:0707.3620 (2007). 
\bibitem[M\"uller et al.(2003)]{muller-biskamp-grappin}  M\"uller, W.-C., 
Biskamp, D., \& Grappin, R., Phys. Rev. E~{\bf 67} (2003) 066302. 
\bibitem[M\"uller \& Grappin(2005)]{muller} M\"uller, W.-C. \& Grappin, R., Phys. Rev. Lett., {\bf 95} (2005) 114502. 
\bibitem[Newell et al.(2001)]{newell} Newell, A. C., Nazarenko, S., \& Biven, L., Physica D {\bf 152-153} (2001) 520.
\bibitem[Ng \& Bhattacharjee(1996)]{ng96} Ng, C. S. \& Bhattacharjee, A., ApJ {\bf 465} (1996) 845.
\bibitem[Ng et al.(2003)]{ng}Ng, C. S., Bhattacharjee, A., Germashewski, K. \& Galtier, S., Phys. Plasmas {\bf 10} (2003) 1954. 
\bibitem[Podesta et al.(2006)]{podesta} Podesta, J. J., Roberts, D. A., \& Goldstein, M. L., J.Geophys. %%@
Res. {\bf 111} (2006) A10109. 
\bibitem[Politano \& Pouquet(1998)]{politano-pouquet} Politano, H. \& Pouquet, A., Geophys. Res. Lett. {\bf 25} (1998) 273.
\bibitem[Rapazzo et al.(2007)]{rapazzo} Rappazo, A. F., Velli, M., Einaudi, G., \& Dalburgh, R. B., ApJ {\bf 657} (2007) L47.
\bibitem[Shebalin et al.(1983)]{Shebalin} Shebalin, J.~V., Matthaeus, W.~H. \& Montgomery, D., J.~Plasma Physics, {\bf 29} (1983) 525.
\bibitem[Smirnova et al.(2006)]{smirnova} Smirnova, T. V., Gwinn, C. R., \& Shishov, V. I., Astron. Astrophys. {\bf 453} (2006) 601. 
\bibitem[Zakharov et al.(1992)]{zakharov} Zakharov, V. E., L'vov, V. S. \& Falkovich, G., {\em Kolmogorov Spectra of Turbulence} (Springer, Berlin,  1992).

\end{thebibliography}
\end{document}